\documentclass[aps,prb,reprint]{revtex4-1}

\usepackage{graphicx}% Include figure files
\usepackage{epstopdf}
\usepackage{dcolumn}% Align table columns on decimal point
\usepackage{bm}% bold math
\usepackage{color}
\usepackage{natbib}
\usepackage{amsmath}
\usepackage{amssymb}
\usepackage[usenames,dvipsnames]{xcolor}

\definecolor{cream}{RGB}{222,217,201}

\begin{document}

\title{`Ferroelectric' Metals Reexamined: Fundamental Mechanisms and Design Considerations for New Materials}

\author{Nicole A. Benedek}
\affiliation{Department of Materials Science and Engineering \\Cornell University, Ithaca, New York 14853, USA}
\email{nbenedek@cornell.edu}
\author{Turan Birol}
\affiliation{Department of Physics and Astronomy \\Rutgers University, Piscataway, New Jersey 08854, USA}
\email{tbirol@physics.rutgers.edu}

\begin{abstract}
The recent observation of a ferroelectric-like structural transition in metallic LiOsO$_3$ has generated a flurry of interest in the properties of polar metals. Such materials are thought to be rare because free electrons screen out the long-range electrostatic forces that favor a polar structure with a dipole moment in every unit cell. In this work, we question whether long-range electrostatic forces are always the most important ingredient in driving polar distortions. We use crystal chemical models, in combination with first-principles Density Functional Theory calculations, to explore the mechanisms of inversion-symmetry breaking in LiOsO$_3$ and both insulating and electron-doped ATiO$_3$ perovskites, A = Ba, Sr, Ca. Although electrostatic forces do play a significant role in driving the polar instability of BaTiO$_3$ (which is suppressed under electron doping), the polar phases of CaTiO$_3$ and LiOsO$_3$ emerge through a mechanism driven by local bonding preferences and this mechanism is `resistant' to the presence of charge carriers. Hence, our results suggest that there is no fundamental incompatibility between metallicity and polar distortions. We use the insights gained from our calculations to suggest design principles for new polar metals and promising avenues for further research.
\end{abstract}

\maketitle

%\begin{abstract}
%The recent observation of a ferroelectric-like structural transition in metallic LiOsO$_3$ has generated a flurry of interest in the properties of polar metals. Such materials are thought to be rare because free electrons screen out the long-range electrostatic forces that favor a polar structure with a dipole moment in every unit cell. In this work, we question whether long-range electrostatic forces are always the most important ingredient in driving polar distortions. We use crystal chemical models, in combination with first-principles Density Functional Theory calculations, to explore the mechanisms of inversion-symmetry breaking in LiOsO$_3$ and both insulating and electron-doped ATiO$_3$ perovskites, A = Ba, Sr, Ca. Although electrostatic forces do play a significant role in driving the polar instability of BaTiO$_3$ (which is suppressed under electron doping), the polar phases of CaTiO$_3$ and LiOsO$_3$ emerge through a mechanism driven by local bonding preferences and this mechanism is `resistant' to the presence of charge carriers. Hence, our results suggest that there is no fundamental incompatibility between metallicity and polar distortions. We use the insights gained from our calculations to suggest design principles for new polar metals and promising avenues for further research.
%\end{abstract}

%%%MAIN TEXT%%%%
\section{Introduction}
\textit{``The question at the root of all crystal chemistry is: Why do the observed structures exist, rather than others we might have thought of with the same chemical composition? Rarely, if at all, can this be answered quantitatively, but qualitatively we can often give very good reasons.''}\cite{megaw73}\\
Writing in 1973, pioneering crystallographer Helen Megaw posed a question that is at the heart of the current materials-by-design effort. Although structure-\emph{property} relationships are usually considered the starting point for materials design, elucidating the fundamental structure-\emph{composition} relationships for a given material very often also results in new insights into the material's properties; these insights can subsequently lead to new design possibilities. Megaw goes on to note that (in her time, at least), crystal chemistry was mostly qualitative, since ``observation is still ahead of theory''.\cite{megaw73} However, the development of powerful first-principles theoretical techniques (such as Density Functional Theory), together with the advent of high-performance supercomputers, means that we can now answer crystal chemical questions quantitatively. In addition, theory is increasingly able to make experimentally testable and realizable predictions. Hence, theory is often an equal partner with ``observation'' (experiment), and sometimes even leads it.

Theory can be an especially powerful tool for understanding and designing materials with purportedly contra-indicated properties. For example, the dearth of ABO$_3$ perovskite multiferroics was ascribed to an incompatibility between the acentric B-site displacements that give rise to ferroelectricity in $d^0$ perovskites such as BaTiO$_3$, and magnetism, which would require a partially filled $d$ shell.\cite{hill00} However, Ref. \onlinecite{hill00} suspected that alternative ferroelectric mechanisms, those for which the driving force is electrostatic interactions rather than charge transfer (as in perovskite titanates), may allow for the co-existence of ferroelectricity and magnetism.\cite{vanaken04,fennie05,ederer06} Subsequent theoretical work by various groups revealed the existence of one such mechanism. In so-called ``trilinear coupling'' or ``hybrid improper''\cite{perezmato04,dawber08,etxebarria10,benedek11,rondinelli12,mulder13} ferroelectrics (that contain magnetic cations), acentric atomic displacements induce not only ferroelectricity, but also ferromagnetism and magnetoelectricity. In the layered perovskite compounds that have been studied so far, the lattice distortions that drive ferroelectricity are actually non-polar `rotations' of the BO$_6$ octahedra, which are generally thought to be driven by local electrostatic, ion size mismatch effects related to the A-site cation bonding environment. The octahedral rotations then couple to a polar mode to produce a non-zero polarization. This mechanism was recently experimentally verified for one of the first predicted hybrid improper ferroelectrics.\cite{oh15} The polarization in a multiferroic double perovskite was also shown to arise through a trilinear coupling mechanism.\cite{pitcher15}

The recent observation\cite{shi13} of a ferroelectric-like structural transition in metallic LiOsO$_3$ has generated a flurry of interest in materials that simultaneously exhibit another pair of supposedly contra-indicated properties: polarity and metallicity. Polar metals are thought to be rare because free electrons screen out the long-range electrostatic forces that favor a polar structure with a dipole moment in every unit cell. However, such materials are of interest, particularly because in many cases they have the potential to exhibit and provide opportunities to explore exotic quantum phenomena.\cite{ncsbook} For example, antisymmetric spin-orbit interactions in polar and non-centrosymmetric superconductors are thought to give rise to non-standard pairing mechanisms and a host of unusual and fascinating properties: extremely large and highly anisotropic upper critical fields, topologically protected spin currents, and complex phase diagrams involving superconductivity and magnetism. Polar metals may facilitate the design of materials with controllable metal-insulator transitions and, paradoxically, \emph{insulating} multiferroics, as shown by recent theoretical predictions on LiOsO$_3$/LiNbO$_3$ superlattices.\cite{puggioni15} Highly conductive ferroelectric oxides -- those with carrier concentrations close to a metal-insulator transition -- are of interest for oxide-based thermoelectrics.\cite{lee12} The practical applications of polar metals and highly conductive ferroelectrics are largely unexplored but promising avenues for further research.

\begin{figure}
\centering
\includegraphics[width=8.0cm]{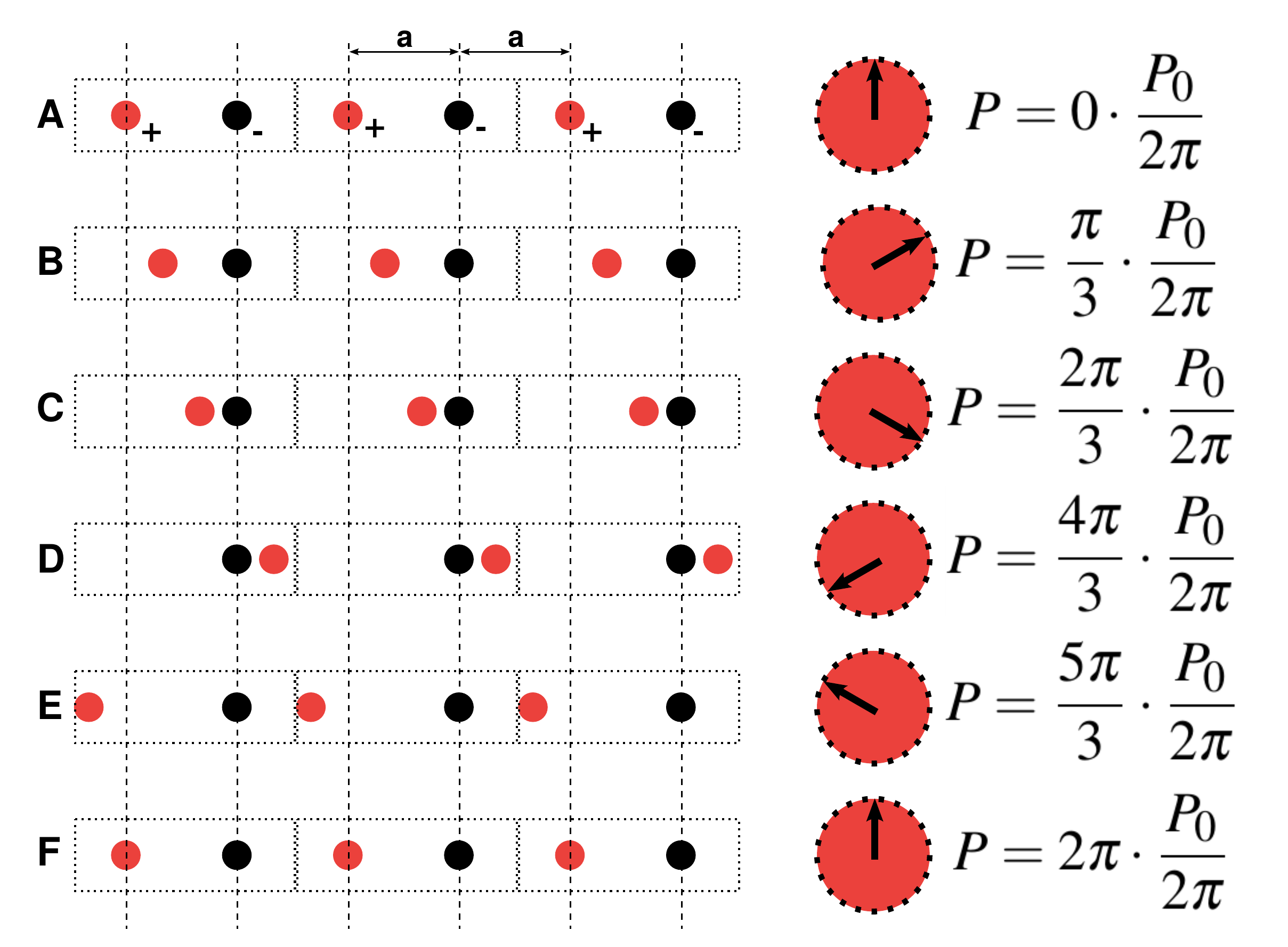}
\caption{\label{fig:PolarizationPhase}
Sketch of a one-dimensional chain of positive and negative point charges. Starting from a centrosymmetric chain in which the positively and negatively charged ions are equidistant (Panel A), which is considered to have zero polarization in the local dipole picture, displacements of the positively charged ion in the positive $x$ direction (Panels B-D) will create a positive polarization. The polarization keeps increasing for even larger values of displacement (panel E) but should go back to zero when the positive atom is displaced by a whole lattice vector (panel F). In the local dipole picture however, panels A and F may have different absolute values of the polarization. In the modern theory of polarization, the polarization is defined only modulo a quantum of polarization, and so the polarizations of $P_0$ and $0$ are equal to each other. This becomes clearer when one considers an angle that is proportional to the value of polarization: In the modern theory, the value of the polarization is meaningful only modulo a quantum of polarization ($P_0$), just like the value of an angle is meaningful only modulo 2$\pi$ (in the sketch, increasing the polarization by $P_0$ by displacing an atom by a unit cell corresponds to this angle increasing by 2$\pi$). The local dipole picture of polarization leads to inconsistencies because in that picture, the absolute value of $P$ (\textit{not} modulo anything) is considered to have a physical meaning.}
\end{figure}

In this paper, we explore the question of whether `ferroelectricity' and metallicity are really contra-indicated and use the insights gained to suggest design principles for new polar metals. As a start, we better define the problem at hand by identifying two separate, but related, issues. First, is it possible for a metal to be an actual ferroelectric? That is, is it possible for a metal to display a spontaneous polarization, the direction of which can be switched with an applied electric field? The second question is, can a metal form in or undergo a phase transition to a polar space group?  It is helpful to initially consider the first of these questions for insulators. In 1950, Slater\cite{slater50} formulated a model of ferroelectricity in which individual unit cell-level dipoles are aligned in the same direction by long-range electrostatic forces. This alignment of dipoles gives rise to a macroscopic polarization, which can be manipulated with electric fields. In the simplest local dipole picture, if there are $n$ atoms in the unit cell of a crystal with positions $\vec{r_i}$ and charges $q_i$, the dipole moment $\vec{D}$ of a unit cell is given by 
\begin{equation}
\vec{D}=\sum_i^n q_i \vec{r_i},
\label{dip} 
\end{equation}
which, when divided by the unit cell volume $V$, gives the polarization
\begin{equation}
\vec{P}=\frac{\vec{D}}{V} .
\end{equation}
The development of the modern theory of polarization\cite{kingsmith93,resta93,vanderbilt93} has since shown that the local dipole picture of ferroelectricity is ``neither a realistic nor useful one'',\cite{resta07} since in a real crystalline material it is impossible to partition the total polarization into localized contributions from unit cell-level dipoles in a non-arbitrary way (see the overview by Spaldin for a beautiful and accessible introduction to the modern theory of polarization from a solid-state chemistry perspective\cite{spaldin12}). %
One of the many problems of the local dipole picture stems from the periodicity of a bulk crystal lattice. Consider the one-dimensional chain of alternating positive and negative ions shown in Figure \ref{fig:PolarizationPhase}. Displacement of a positively charged atom by a lattice vector gives an arrangement of the chain that is indistinguishable from that before the displacement. However, we have moved a charge by the length of the unit cell (or equivalently, by one lattice vector) and so according to Equation \ref{dip}, we have generated a dipole moment. We can move the positively charged atom by any integer number of lattice vectors, and each time we will generate a different dipole moment! Hence, in the local dipole picture, in which the absolute value of the polarization is a meaningful quantity, physically equivalent configurations of atoms can give rise to physically different polarizations. 
The modern theory of polarization does away with local dipoles, and instead recasts the problem in terms of a reciprocal space quantity, a particular \textit{phase} of the occupied Bloch states. 
The absolute value of the polarization in this theory, being related to a phase, is arbitrary modulo a \textit{polarization quantum}. In the context of Figure \ref{fig:PolarizationPhase}, a polarization quantum is the value of polarization resulting from the displacement of the positively charged ion by one unit cell.  
Even though the absolute value of the polarization is multi-valued, the quantity that is actually measured experimentally -- the \textit{change} in polarization under electric field switching -- is single valued. The local dipole picture assigns a physical meaning to the absolute value of the polarization, whereas the modern theory of polarization recognizes that the absolute values of the polarization for a given material are all equivalent, being related by the polarization quantum. In the modern theory of polarization, only the change in polarization under electric field switching is physically meaningful. The polarization of an infinite insulator is defined as a bulk property, the magnitude of which depends on the details of the crystal structure and bonding of a given material. This is in contrast to metals, where the modern theory cannot be used to even define a polarization. This approach is applicable only when there is a gap between the occupied and unoccupied energy levels (it is even possible to come up with a rigorous definition of a metal as a system for which a polarization cannot be defined in bulk\cite{resta02}). In this respect then, a metal cannot be said to be a ferroelectric.

Coming now to the second issue, having said the above, the question of whether the polarization can be rigorously defined for a metal is quite distinct from that of whether it can undergo a phase transition to or form in a polar space group. We will use the terms polar metal and `ferroelectric' metal interchangeably to describe materials that belong to one of the 10 polar crystal classes and have a non-zero density of states at the Fermi level. The usual driving force for polar distortions is assumed to be long-range electrostatic forces. Is this picture correct? This is the question we investigate in this paper. First, we present data from a search of the Inorganic Crystal Structure Database (ICSD) showing that there are a not insignificant number of `ferroelectric' or polar metals. Given the importance of interdisciplinary interactions between solid-state chemists and physicists and the fact that polar materials are of interest to both communities, we then explore the interaction between `polarity' and metallicity in detail from two different perspectives.  We first use crystal chemical models and first-principles density functional theory calculations to investigate the mechanism of inversion symmetry-breaking in the recently synthesized polar metal LiOsO$_3$. We then consider the behavior of the polar instability in electron-doped BaTiO$_3$, SrTiO$_3$ and CaTiO$_3$ from the perspective of lattice dynamics. Our results appear to indicate that long-range electrostatic forces may be of secondary importance for the emergence of polar structures in metals, and suggest that there is no fundamental incompatibility between between polarity and metallicity. Although it may appear as if we are primarily concerned with microscopic mechanisms of inversion-symmetry breaking in metals, understanding the fundamental interactions between different physical/chemical effects or functional properties, \emph{and} connecting that information to crystal chemistry, lies at the foundation of materials design. It is difficult to formulate reliable structure-property relationships if the fundamentals of a given property are not clear or fully understood. Hence, we use the insights gained from our investigation of LiOsO$_3$ and the titanate perovskites to suggest design principles for new polar metals.

\section{Computational Details}
Our first-principles calculations were performed using density functional theory with projector augmented wave potentials,\cite{PAW1, PAW2} as implemented in the Vienna ab initio Simulation Package (VASP).\cite{VASP1, VASP2, VASP3}  Calculations on LiOsO$_3$ were performed using the PBEsol exchange-correlation functional, a plane wave cutoff of 600 eV and a 12$\times$12$\times$12 \textbf{k}-point mesh. We applied a Gaussian smearing of 0.1 eV and ignored the effects of spin-orbit coupling, since the Os $d$ bands have quite a large width. In addition, even though spin-orbit coupling can give rise to many interesting magnetic and topological properties, it generally does not strongly affect crystal structure. Since there is no evidence from experiment\cite{shi13} that LiOsO$_3$ displays any long-range magnetic order, our calculations were performed with Os in a non-magnetic configuration. A force-convergence tolerance of 2.5 meV/\AA~was used for all structural relaxations and lattice dynamical properties (phonon frequencies and eigenmodes) were calculated using density functional perturbation theory.

For the titanates, we relaxed the cubic perovskite structure for each material also with the PBEsol functional and used these relaxed structures (3.847 \AA~for CaTiO$_3$ and 3.985 \AA~for BaTiO$_3$) for subsequent phonon calculations. The phonon calculations were performed using the direct method and the displacements shown in Figure \ref{fig:phonons} as a basis. No acoustic sum rule or symmetrization of the force constant matrices was applied since the force constants matrices obtained carried the correct symmetries with only a small numerical error. The spin-orbit interaction in 3$d$ transition metals is of the order of few tens of meV, which is much smaller than the energy scales relevant to the present problem, and on this basis, we ignored it. We used a plane wave cutoff of 500 eV and the equivalent of a $8\times 8\times 8$ $\Gamma$ centered \textbf{k}-point mesh for Brillouin zone integrals.

\section{Results and Discussion}
\subsection{A Brief History of `Ferroelectric' Metals}
As a prelude to the presentation and discussion of results, some brief historical context for the study of `ferroelectric' metals is given. Anderson and Blount\cite{anderson65} first suggested the possibility of a ferroelectric-like structural transition in a metal in their study of structural phase transitions in V$_3$Si. This material (and many others in the same family) is a superconductor that adopts the cubic A15 or $\beta$-W structure above $\sim$21 K, below which it undergoes a transition to a tetragonal phase. Experiments had shown that the transition was second order and it was described as ``martensitic'', in the sense that no diffusion was involved. The designation was somewhat confusing to researchers at the time because most of the known martensitic phase transitions were strongly first order and characterized by a distortion involving a change in the shape of the unit cell only, \textit{i.e.} a strain. Anderson and Blount used Landau theory to show that it was not possible to describe the transition in V$_3$Si using strain as the only order parameter. They thus concluded that if the experimental observations were correct and the phase transition really was second order, then some unknown order parameters other than strain must be involved. It was then suggested that the simplest explanation as to the unknown order parameters was that they were associated with atomic displacements that globally break inversion symmetry in the tetragonal phase. Hence, Anderson and Blount's focus was on rationalizing the critical behavior of V$_3$Si, rather than on making predictions about or understanding the physics of `ferroelectric' metals.

As we will see in the next section, the ICSD contains a number of entries for `ferroelectric' metals, suggesting that these kinds of materials are not especially rare. However, almost all of the `ferroelectric' metals in the ICSD crystallize in a polar space group and thus never undergo a transition from a non-polar to a polar structure. In discussing the critical behavior of V$_3$Si, Anderson and Blount discuss second order phase transitions in the context of BaTiO$_3$, the best known ferroelectric at the time, which undergoes such a transition. Anderson and Blount's discussion of ferroelectricity in BaTiO$_3$ is often interpreted as listing conditions that a material must fulfil in order to be considered a `ferroelectric-like' metal. One of these conditions is that the material must undergo a continuous transition from the non-polar to the polar phase, although such a transition will not be important to the properties of the metal in the polar state. 

Later studies on V$_3$Si cast doubt on the initial reports that the phase transition is second order and that the low temperature structure is polar.\cite{testardi75,brown01} The pyrochlore Cd$_2$Re$_2$O$_7$ was initially classified as a `ferroelectric' metal,\cite{sergienko04} however although the lowest-symmetry phase is non-centrosymmetric, it is non-polar.\cite{ishibashi10} The first unambiguous report of a `ferroelectric' metal -- that is, a metal that undergoes a continuous phase transition to a polar structure -- appeared only very recently, in 2013.\cite{shi13} Shi and co-workers used high-pressure techniques to synthesize a new perovskite-like material, LiOsO$_3$, and showed that it undergoes a continuous transition at 140 K from the centrosymmetric space group $R\bar{3}c$ to the non-centrosymmetric and polar space group $R3c$. Resistivity measurements confirmed that LiOsO$_3$ is metallic in both phases, though the resistivity in the polar phase is more than an order of magnitude greater than a normal metal. Neutron diffraction data show that the transition is accompanied by a shifting of the Li ions along the cubic perovskites [111] axis, in exactly the same manner as LiNbO$_3$ and LiTaO$_3$; subsequent first-principles calculations\cite{capone14,xiang14,sim14,liu15} confirmed the role of Li displacements in the transition to the polar phase. The transition mechanism will be discussed in detail in Section \ref{Li-sec}.

\subsection{Crystallographic Survey of `Ferroelectric' Metals}
Table \ref{icsd_data} shows the results of a search of the ICSD for `ferroelectric' metals. The search was restricted to metals in polar space groups only (non-centrosymmetric but non-polar metals were not included) and was not exhaustive. A number of known non-centrosymmetric superconductors are not listed, nor are most of the many known intermetallics that form in polar space groups. Nonetheless, Table \ref{icsd_data} contains approximately 70 entries. Before continuing our discussion, we note that less than 10\% of all materials in the ICSD are polar\cite{bennett12} and so, in a relative sense, polar materials in general can be considered rare. However, there are approximately 20,000 known intermetallics,\cite{dshemuchadse15,dshemuchadse15b} taking one class of materials as an example, and if we assume that 10\% of them form in polar space groups, we are still left with a significant number of compounds (even accounting for the fact that not all intermetallics are metals). Hence, in absolute terms polar metals are not generally rare, though they do appear to be scarce in certain families of materials. For example, Table \ref{icsd_data} lists just two complex oxides: LiOsO$_3$\cite{shi13} and the Ruddlesden-Popper phase Ca$_3$Ru$_2$O$_7$.\cite{yoshida05} A more thorough exploration, focused on oxides, was conducted by searching for metals among the 388 known (up to 1998) polar oxides listed in Ref. \onlinecite{halasyamani98}. Of these materials, many could be immediately discounted based on chemistry and stoichiometry considerations; physical properties data could not be found for most of the remaining materials. Just three materials of the 388 could be confirmed as metals: Na$_{0.9}$Mo$_6$O$_{17}$,\cite{onoda86} Na$_{0.6}$CoO$_2$,\cite{fouassier73,greenblatt85,yang04} and Bi$_{10}$Sr$_{10}$Cu$_5$O$_{29}$.\cite{onoda88} This should not surprise us too much, considering that the chemical compositions of most oxides result in electronic structures that are insulating. In addition, the crystal chemistry of some large families of complex oxides, such as the perovskites, favors non-polar structural distortions, as discussed in Ref. \onlinecite{benedek13}. Hence, there is nothing particularly `special' about oxides (compared to any other class of materials) that prevents them from being either polar or metallic, it is simply that the chemical compositions of most oxides result in electronic structures that are insulating, and crystal chemistries that give rise to non-polar, instead of polar, structures. More generally, our crystallographic survey suggests that there is no fundamental contra-indication between polarity and metallicity, since one would not expect to find so many materials simultaneously exhibiting both properties if they were truly incompatible. In the following sections, we investigate the interaction between polarity and metallicity at the microscopic level for two model systems, LiOsO$_3$ and ATiO$_3$ perovskites (A = Ca, Sr, Ba).

\begin{table}
\small
\caption{Survey of a selection of polar metals in the Inorganic Crystal Structure Database classified according to crystal class and space group}
\label{icsd_data}
\begin{tabular*}{0.5\textwidth}{@{\extracolsep{\fill}}lcc}
\hline
Material&Space Group&Reference\\
\hline
\textbf{6mm}\\
CeAuGe&$P6_3mc$&\onlinecite{pottgen96}\\
LuAuGe&$P6_3mc$&\onlinecite{pottgen96b}\\
ScAuGe&$P6_3mc$&\onlinecite{pottgen96b}\\
HoAuGe&$P6_3mc$&\onlinecite{gibson01}\\
CeCuSn&$P6_3mc$&\onlinecite{maehlen05}\\
La$_{15}$Ge$_9$C&$P6_3mc$&\onlinecite{wrubl11}\\
La$_{15}$Ge$_9$Fe&$P6_3mc$&\onlinecite{guloy96}\\
La$_{15}$Ge$_9$Co&$P6_3mc$&\onlinecite{guloy96}\\
La$_{15}$Ge$_9$Ni&$P6_3mc$&\onlinecite{guloy96}\\
Sr$_3$Cu$_8$Sn$_4$&$P6_3mc$&\onlinecite{pani11}\\
IrMg$_{2.03}$In$_{0.97}$&$P6_3mc$&\onlinecite{hlukhyy04}\\
IrMg$_{2.20}$In$_{0.80}$&$P6_3mc$&\onlinecite{hlukhyy04}\\
\textbf{6}\\
CaAlSi&$P6_3$&\onlinecite{sagayama06}\\
TlV$_6$S$_8$&$P6_3$&\onlinecite{ohtani01}\\
KV$_6$S$_8$&$P6_3$&\onlinecite{ohtani01}\\
RbV$_6$S$_8$&$P6_3$&\onlinecite{ohtani01}\\
CsV$_6$S$_8$&$P6_3$&\onlinecite{ohtani01}\\
\textbf{4mm}\\
REPt$_3$B, RE = La, Pr, Nd&$P4mm$&\onlinecite{sologub03}\\
LaRhSi$_3$&$I4mm$&\onlinecite{lejay84}\\
LaIrSi$_3$&$I4mm$&\onlinecite{lejay84}\\
RECoGe$_3$, RE = Ce, La&$I4mm$&\onlinecite{pecharsky93}\\
CeRhGe$_3$&$I4mm$&\onlinecite{kawai08}\\
CeRuSi$_3$&$I4mm$&\onlinecite{kawai08}\\
LaIrGe$_3$&$I4mm$&\onlinecite{kawai08b}\\
LaFeGe$_3$&$I4mm$&\onlinecite{kawai08b}\\
LaRh$_3$&$I4mm$&\onlinecite{kawai08b}\\
PrCoGe$_3$&$I4mm$&\onlinecite{kawai08b}\\
CaIrSi$_3$&$I4mm$&\onlinecite{eguchi10}\\
CaPtSi$_3$&$I4mm$&\onlinecite{eguchi11}\\
SrAuSi$_3$&$I4mm$&\onlinecite{isobe14}\\
EuPdGe$_3$&$I4mm$&\onlinecite{kaczorowski12}\\
EuPtSi$_3$&$I4mm$&\onlinecite{kumar10}\\
REPdIn$_2$, RE = Pr, Nd, Sm, Gd, Er, Tm, Lu&$I4mm$&\onlinecite{zaremba04}\\
La$_2$NiAl$_7$&$I4mm$&\onlinecite{gout05}\\
SnP&$I4mm$&\onlinecite{donohue70}\\
GeP&$I4mm$&\onlinecite{donohue70b}\\
\textbf{3mm}\\
Ir$_9$Al$_28$&$P31c$&\onlinecite{katrych06}\\
$\gamma$-Bi$_2$Pt&$P31m$&\onlinecite{kaiser14}\\
Au$_{6.05}$Zn$_{12.51}$&$P31m$&\onlinecite{thimmaiah13}\\
Ba$_{21}$Al$_{40}$&$P31m$&\onlinecite{amerioun04}\\
Li$_{17}$Ag$_3$Sn$_6$&$P31m$&\onlinecite{lupu04}\\
Cr$_5$Al$_8$&$R3m$&\onlinecite{brandon74}\\
Mn$_5$Al$_8$&$R3m$&\onlinecite{ellner90}\\
Cu$_{7.8}$Al$_5$&$R3m$&\onlinecite{kisi91}\\
Cu$_7$Hg$_6$&$R3m$&\onlinecite{lindahl69}\\
NbS$_2$&$R3m$&\onlinecite{morosin74}\\
Pr$_2$Fe$_{17}$&$R3m$&\onlinecite{calestani03}\\
Pr$_2$Co$_{17}$&$R3m$&\onlinecite{calestani03}\\
Sn$_4$As$_3$&$R3m$&\onlinecite{eckerlin68}\\
Sn$_4$P$_3$&$R3m$&\onlinecite{eckerlin68}\\
LiOsO$_3$&$R3c$&\onlinecite{shi13}\\
\textbf{mm2}\\
La$_4$Mg$_5$Ge$_6$&$Cmc2_1$&\onlinecite{solokha12}\\
La$_4$Mg$_7$Ge$_6$&$Cmc2_1$&\onlinecite{solokha12}\\
Ca$_3$Ru$_3$O$_7$&$Cmc2_1$&\onlinecite{yoshida05}\\
Yb$_2$Ga$_4$Ge$_6$&$Cmc2_1$&\onlinecite{zhuravleva04}\\
Ce$_2$Rh$_3$(Pb, Bi)$_5$&$Cmc2_1$&\onlinecite{thomas07}\\
Eu$_2$Pt$_3$Sn$_5$&$Cmc2_1$&\onlinecite{harmening09}\\
Lu$_4$Zn$_5$Ge$_6$&$Cmc2_1$&\onlinecite{reker13}
\end{tabular*}
\end{table}

\subsection{\label{Li-sec}Fundamental Mechanism Driving Polarity in LiOsO$_3$: Crystal Chemical Perspective}
Shi and co-workers\cite{shi13} used neutron diffraction to show that the $R\bar{3}c$ to $R3c$ transition in LiOsO$_3$ is accompanied by a shift of about 0.5 \AA~in the mean positions of the Li atoms along the cubic perovskite [111] axis (equivalent to the $c$ axis in the $R\bar{3}c$ and $R3c$ space groups in the hexagonal setting). Subsequent first-principles calculations\cite{sim14,xiang14,capone14,liu15} showed that the centrosymmetric $R\bar{3}c$ phase is unstable to a zone-center phonon that breaks inversion symmetry and produces the $R3c$ space group. These calculations also showed that this mode is dominated by Li displacements, in agreement with experimental observations. The ferroelectric mechanism in LiOsO$_3$ in fact appears very similar to that identified previously\cite{benedek13} for $R3c$ materials, such as FeTiO$_3$ and ZnSnO$_3$, which are isostructural with LiOsO$_3$, LiNbO$_3$ and LiTaO$_3$. In these materials, the coordination environment of the (very small) A-site cations is optimized through a combination of rotations of the BO$_6$ octahedra, which are non-polar, and a polar displacement of the A-site cations. Neither distortion is accompanied by any significant changes in charge transfer or hybridization. Is this how a polar structure emerges in LiOsO$_3$?

\begin{figure}
\centering
\includegraphics[width=9.5cm]{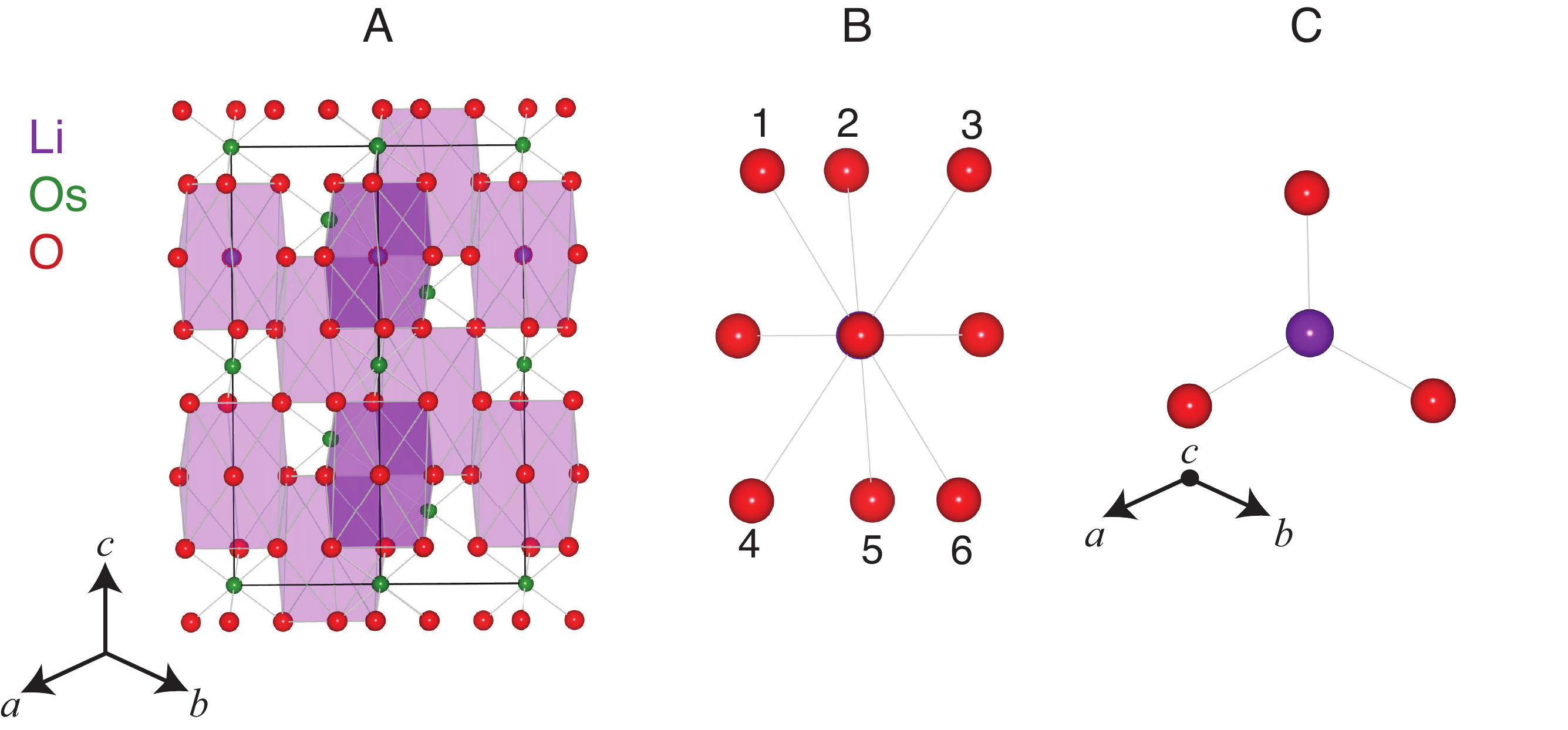}
\caption{\label{Li_coord} Structure of LiOsO$_3$ in the non-polar R$\bar{3}c$ phase (hexagonal setting) with the Li coordination polyhedra highlighted. The six long bonds to Li are labeled in panel B. The unlabeled oxygen ions form the three short bonds, shown in panel C. Note that panels A and B share the same axes (shown at left), whereas panel C is being viewed from a different direction. See also Figure 11 of Ref. \cite{benedek13}.}
\end{figure}

Before we begin exploring the mechanism through which LiOsO$_3$ transitions from $R\bar{3}c$ to $R3c$, we pause to make a few comments regarding our analysis below. One of the simplest and most insightful models for studying the crystal chemistry of structural distortions is the bond valence model. Since the bond valence model is essentially based on Pauling's rules (which were developed to rationalize the structures of ionic crystals), it is not generally applicable to metallic systems, or those that exhibit strong electronic correlations. There is no conclusive evidence from either theory or experiment that LiOsO$_3$ exhibits significant electronic correlations,\footnote{Some signatures of correlations were observed in infrared spectroscopy and first-principles calculations in a recent study,\cite{vecchio15} however transport and calorimetry measurements suggest that LiOsO$_3$ is an uncorrelated metal.\cite{shi13}} however it is obviously a metal. Nonetheless, we will use the bond valence model to elucidate the ferroelectric mechanism of LiOsO$_3$, an approach we believe is justified (at least on qualitative grounds), for two reasons. First, even though LiOsO$_3$ is metallic, it is not unreasonable to assume that the bonding preferences of the constituent atoms are approximately similar to those in similar insulating compounds, such as LiNbO$_3$ or ZnSnO$_3$, for example. Secondly, previous theory\cite{sim14,capone14} has shown that the density of states around the Fermi level is dominated by the Os $d$ states, as one would expect. The majority of the density of states for both Li and O is several eV below the Fermi level, though there is some O character to the density of states around the Fermi level, which probably occurs through mixing with the Os $d$ states. In other words, the Os $d$ states are delocalized, whereas the Li and O $p$ states are more localized (as they would be in an insulating compound) and are therefore more amenable to a bond valence analysis. LiOsO$_3$ is still however ultimately a metal, and so the bond valences quoted for Li below should be interpreted solely as approximate indicators of the ideality of the Li coordination environment, rather than as ionic charges.

\begin{figure}
\centering
\includegraphics[width=6cm]{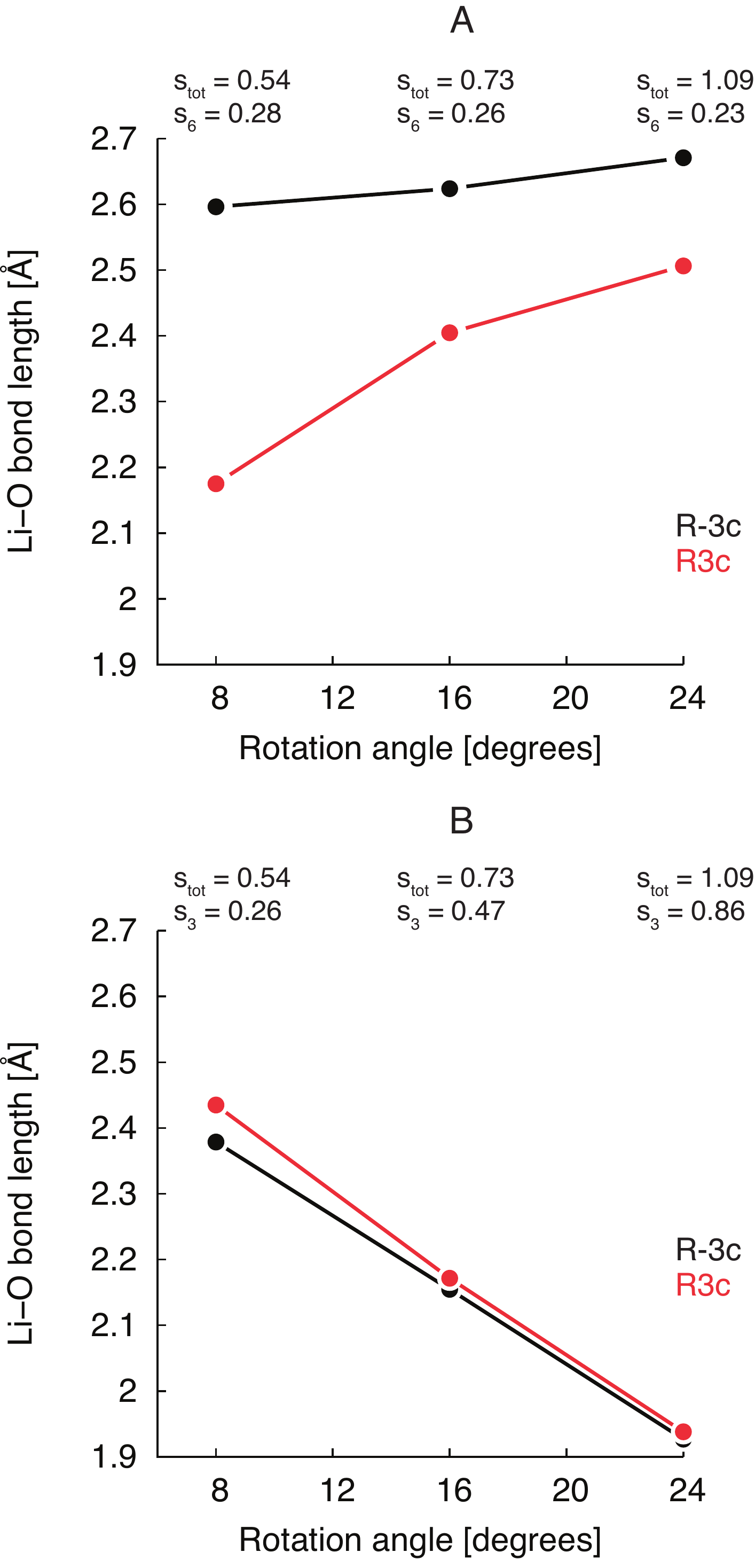}
\caption{\label{Li_bonds}Change in length of Li-O bonds with increasing octahedral rotation angle and polar distortion to $R3c$. Panel A (top) shows the length of the six long bonds (bonds labeled 1-6 in Figure \ref{Li_coord}B) in $R\bar{3}c$ and the length of three of these bonds (1-3) in $R3c$. Panel B (bottom) shows the change in length of the three short Li-O bonds. Both plots are drawn to the same scale. The total bond valence, s$_{tot}$, of the Li atom in $R\bar{3}c$ as a function of rotation angle is shown at the top of each plot in valence units, s$_6$ represents the bond valence due to the six long Li-O bonds (A) and s$_3$ represents the bond valence due to the three short Li-O bonds (B). The lines are guides for the eye.}
\end{figure}

We first examine how octahedral rotations change the A-site coordination environment in $R\bar{3}c$. The A-site is nine-coordinate in $R\bar{3}c$, with six long bonds and three short bonds to oxygen atoms, the latter of which are co-planar with the A-site; see Figure \ref{Li_coord}. In the materials studied in Ref. \onlinecite{benedek13}, the six long A-O bonds increase slightly as the octahedral rotations increase, whereas the three short A-O bonds decrease significantly. Figure \ref{Li_bonds} shows that the same trend can be observed in LiOsO$_3$. Figure \ref{Li_bonds} also shows that at low rotation angles, the Li atom is severely underbonded, as quantified through a bond valence analysis, with the total bond valence distributed almost evenly among the six long and three short bonds. As the rotation angle increases, the total bond valence of the Li atom increases, as expected. However, the fraction attributed to the six long Li-O bonds decreases only very slightly with rotation angle (consistent with those bonds increasing only slightly in length), while an increasing fraction is `transferred' to the three short Li-O bonds, consistent with the sharp decrease in the length of these bonds. In fact, when the rotation angle equals its value in the experimentally determined structure,\cite{shi13} 24$^\circ$, the three short Li-O bonds account for nearly 80\% of the bonding around Li.

How does the polar distortion change the environment around Li? Figure \ref{Li_bonds} shows that the polar distortion lengthens the three short Li-O bonds, though not very much. The six long Li-O bonds have equal lengths in $R\bar{3}c$ (black data points in Figure \ref{Li_bonds}A). The polar distortion significantly shortens three of these bonds (red data points in Figure \ref{Li_bonds}A), while lengthening the other three. These bond length changes reduce the coordination of the Li atom from nine in $R\bar{3}c$ to six in $R3c$: the oxygen atoms forming the three short Li-O bonds, together with those numbered 1-3 in Figure \ref{Li_coord}C, form an octahedral coordination environment around Li. Hence, in $R3c$ both the Li atom and Os atom are octahedrally coordinated. Interestingly, there appears to be a crossover in the contribution of each structural distortion -- octahedral rotations and polar displacements -- to the optimization of the Li bonding environment. When the rotation angle is small, the Li atom is underbonded and the polar displacement significantly improves its coordination environment. However, when the rotations become large the Li atom becomes \emph{overbonded} and the polar distortion actually reduces its bond valence. The bond valence analysis also indicates that it is the octahedral rotations that contribute the most overall to the optimization of the Li coordination environment, since even without a polar distortion, Li almost has its optimal valence at large rotation angles. Further details concerning our calculations, including comparisons between the experimental and our calculated structures for the $R\bar{3}c$ and $R3c$ phases, are provided in the Supporting Information.

\begin{figure}
\centering
\includegraphics[width=6cm]{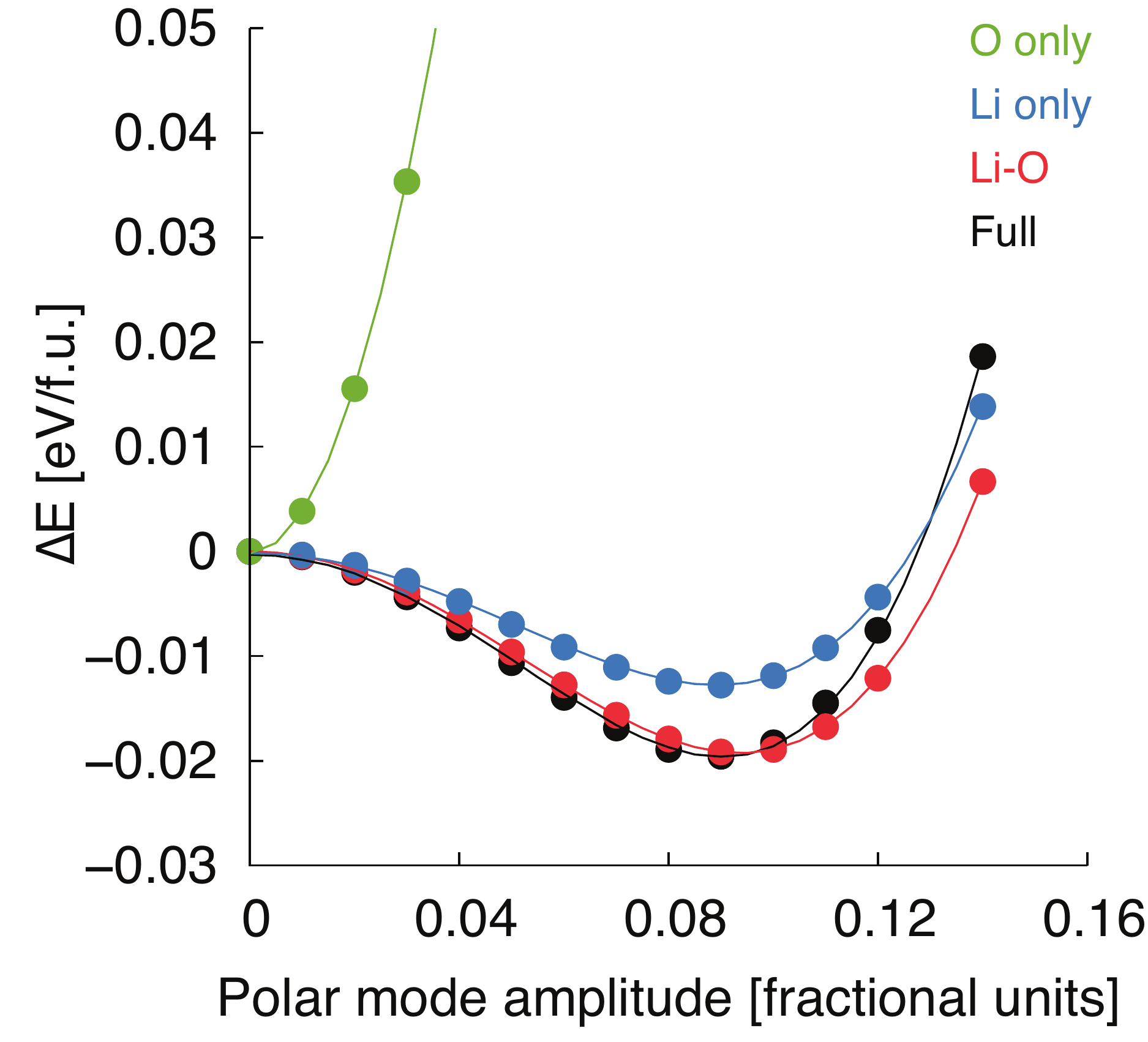}
\caption{\label{fits}Potential energy surface about $R\bar{3}c$ LiOsO$_3$ with respect to the polar distortion. The curves represent displacements of different atoms (or sets of atoms) along the polar mode coordinate. The zero of energy is set to the total energy of the $R\bar{3}c$ phase from first-principles calculations. The lines are fits to functions of the form E = E$_0$ +$\alpha Q^2 + \beta Q^4$, where E$_0$ is the energy without any polar displacements and $Q$ is the amplitude of the polar distortion (terms up to eighth order in $Q$ were included for fits to the Li-O data points only).}
\end{figure}

Our results above suggest that the mechanism through which LiOsO$_3$ undergoes a polar distortion is very similar to that described in Ref. \onlinecite{benedek13}. For both LiOsO$_3$ and the previously studied materials, the coordination preferences of the A-site play a significant role in driving the polar distortion. This appears to be in contrast to LiNbO$_3$ and LiTaO$_3$, for which Inbar and Cohen argue that displacements of the oxygen ions play the dominant role in the transition to the polar phase, and that the Li atoms are ``passive players in the ferroelectric energetics".\cite{inbar96} We investigated the energetics of the polar distortion in LiOsO$_3$ by selectively freezing in the contributions of different sets of atoms to the polar eigenmode, \textit{e.g.} Li atoms only, Li and O together, \textit{etc}. Figure \ref{fits} shows that displacing just the Li and O atoms together while keeping the Os atoms fixed (red data points) lowers the energy almost as much as the full polar distortion (black data points). Indeed, Figure \ref{fits} shows that the Li atom is unstable in $R\bar{3}c$ and can lower the energy significantly by itself (blue data points). Displacing just the O atoms by themselves (green data points) raises the energy, however (displacing just the Os atoms also raises the energy, even more quickly than the O atoms; data not shown). In LiNbO$_3$ and LiTaO$_3$, displacing just the Li atoms by themselves barely lowers the energy at all, whereas displacing just the O atoms does lower the energy. This energy gain is the result of hybridization between the O $p$ states with the Nb/Ta 4$d$/5$d$ states, an interaction which is obviously absent in LiOsO$_3$. Hence, as also argued by Ref. \onlinecite{xiang14}, in terms of the `ferroelectric' mechanism, LiOsO$_3$ appears to have more in common with materials like ZnSnO$_3$ than LiNbO$_3$ and LiTaO$_3$.        

Xiang\cite{xiang14} considers the question of how the Li atoms ``interact with each other'' such that all displacements are in the same direction and result in a polar structure, \textit{i.e.} disorder in the Li displacements would produce an overall non-polar structure. In other words, what is it that compels the Li atoms to align in the same direction in the absence of long-range electrostatic forces? To explore this question, Xiang constructed an effective Hamiltonian\cite{zhong94,zhong95} that included terms describing short-range repulsion and covalency effects. First-principles calculations were used to fit the model parameters. Using this model in Monte Carlo simulations, Xiang was able to reproduce the ferroelectric ground state, the second-order nature of the phase transition and the behavior of the specific heat with temperature. These results suggest that long-range electrostatic forces -- not included in Xiang's model -- are perhaps not as important as one may assume in driving the $R\bar{3}c$ to $R3c$ transition in LiOsO$_3$.

\begin{figure}
\centering
\includegraphics[width=8cm]{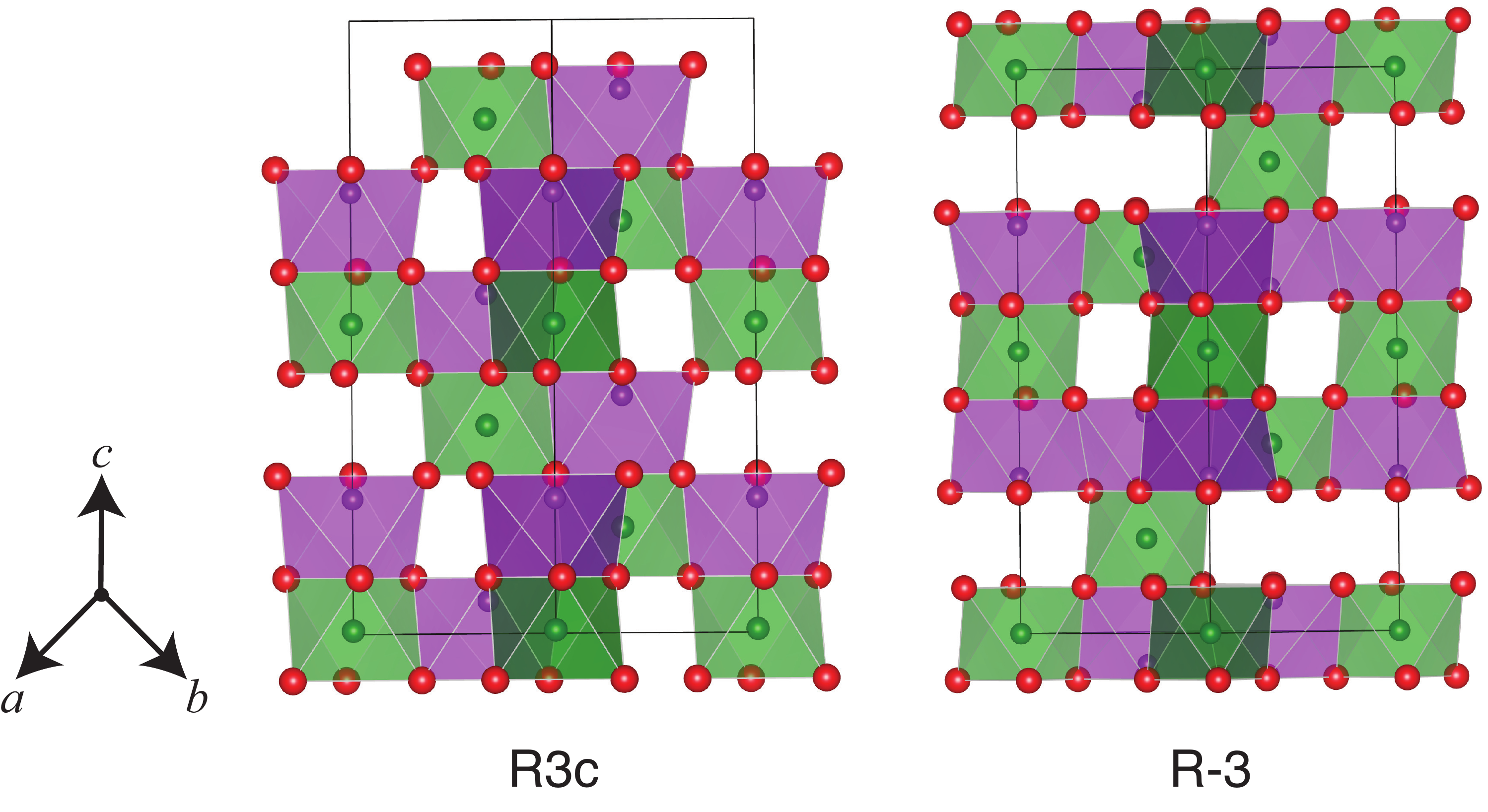}
\caption{\label{R3cR-3} Structure of LiOsO$_3$ in the polar $R3c$ and non-polar $R\bar{3}$ phases. Both structures are shown in the hexagonal setting and the coloring scheme for the atoms is the same as that in Figure \ref{Li_coord}.}
\end{figure}

Instead of long-range electrostatic forces, Xiang instead argues that local bonding requirements favor the parallel alignment of Li atoms. We have already discussed the changes in local bonding that accompany the polar distortion in LiOsO$_3$ -- how would this picture change if instead of aligning in parallel, the Li atoms adopted an \emph{anti-parallel} arrangement? In fact, in addition to the zone-center phonon instability responsible for the polar distortion (in our calculations, this mode has a frequency of $i$148 cm$^{-1}$, in good agreement with previous theory\cite{xiang14}), there is another unstable mode at higher frequency ($i$98 cm$^{-1}$). This mode transforms like the irreducible representation $\Gamma_2^+$ and produces the non-polar space group $R\bar{3}$, which is 0.02 eV/f.u. higher in energy than $R3c$. In contrast to the polar mode discussed above, the $\Gamma_2^+$ mode is characterized by anti-parallel displacements of the Li atoms. Figure \ref{R3cR-3} shows that in $R3c$ pairs of Li and Os-centered octahedra share faces along the hexagonal $c$ axis. The anti-parallel movement of Li atoms along the $c$ axis in $R\bar{3}$ results in an Os-centered octahedron sharing \emph{two} faces with Li-centered octahedra, plus an isolated Os-centered octahedron. We can rationalize the greater stability of the $R3c$ phase in terms of Pauling's third rule, which states that the stability of a structure decreases when polyhedra share edges and, in particular, faces. In an ionic crystal, face-sharing forces the cations in the center of the octahedra to be in closer proximity than they would be if the polyhedra shared only edges or corners. Electronic screening in a metal would mitigate the destabilizing effects of face-sharing polyhedra somewhat, but considering the two polyhedral configurations in the $R3c$ and $R\bar{3}$ phases, it should not be too surprising that having all Li atoms aligned in the same direction is favored over an anti-parallel alignment. Hence, in LiOsO$_3$ at least, it is possible to explain the mechanism through which the polar structure emerges without having to invoke long-range electrostatic forces at all. In the next section, we explore the effects of metallicity on ferroelectricity in a more quantitative manner by considering one of the most well-studied families of ferroelectrics, the titanate perovskites. 

\subsection{Ferroelectricity in Electron Doped Titanate Perovskites: Lattice Dynamical Perspective} 
The first ferroelectrics to be discovered were all hydrogen-containing materials (for example, Rochelle salt NaKC$_4$H$_4$O$_6\cdot$4H$_2$O\cite{valasek21} and KH$_2$PO$_4$\cite{busch35}) and so hydrogen bonding was thought to be a necessary pre-requisite for the existence of ferroelectricity. The discovery of ferroelectricity in perovskite BaTiO$_3$,\cite{wul46} a structurally simple inorganic material, allowed researchers to develop a fundamental understanding of the origin of the phenomenon and to abandon the ``hydrogen hypothesis''.\cite{lines77} Although prior studies\cite{raman40,frohlick49,ginzburg49,ginzburg49b,landau54,anderson60} had suggested a link between soft phonon modes and structural phase transitions, it was Cochran who first pointed out that second order ferroelectric transitions are driven by a Brillouin zone center transverse optical lattice vibration -- the soft mode -- the frequency of which goes to zero as the transition is approached.\cite{cochran59,cochran60} Cochran rationalized this phenomenon by positing that the soft mode frequency is proportional to (the square root of) the difference between the short-range repulsive forces, which favor the non-polar, paraelectric structure, and long-range Coulomb forces, which favor a polar structure with dipole moments in every unit cell. Hence, the soft mode frequency goes to zero and a ferroelectric transition occurs as the difference between the short-range and long-range interactions approaches zero. Although the soft mode concept provided the first microscopic picture of ferroelectricity, it gave no hints as to the chemical driving force for the transition. Later studies revealed that the transition can be considered the result of a pseudo- or Second Order Jahn-Teller (SOJT) distortion,\cite{bersuker66,bersuker78,burdett81,polinger15} driven by hybridization between the Ti 3$d$ and O 2$p$ states.\cite{cohen92,ghosez96} This charge transfer interaction weakens the short-range repulsive forces that favor the cubic structure and allows the long-range Coulomb force to dominate, resulting in a ferroelectric distortion; first-principles calculations have provided many further details on the nature of the polar phase.\cite{yu95, ghosez96, ghosez98, lasota97, ghosez99, rabe07, birol11} In this section, we challenge the assumption that the long-range Coulomb force is always the most important ingredient required for the emergence of a polar structure.

\begin{figure}[h]
\centering
  \includegraphics[width=0.45\textwidth]{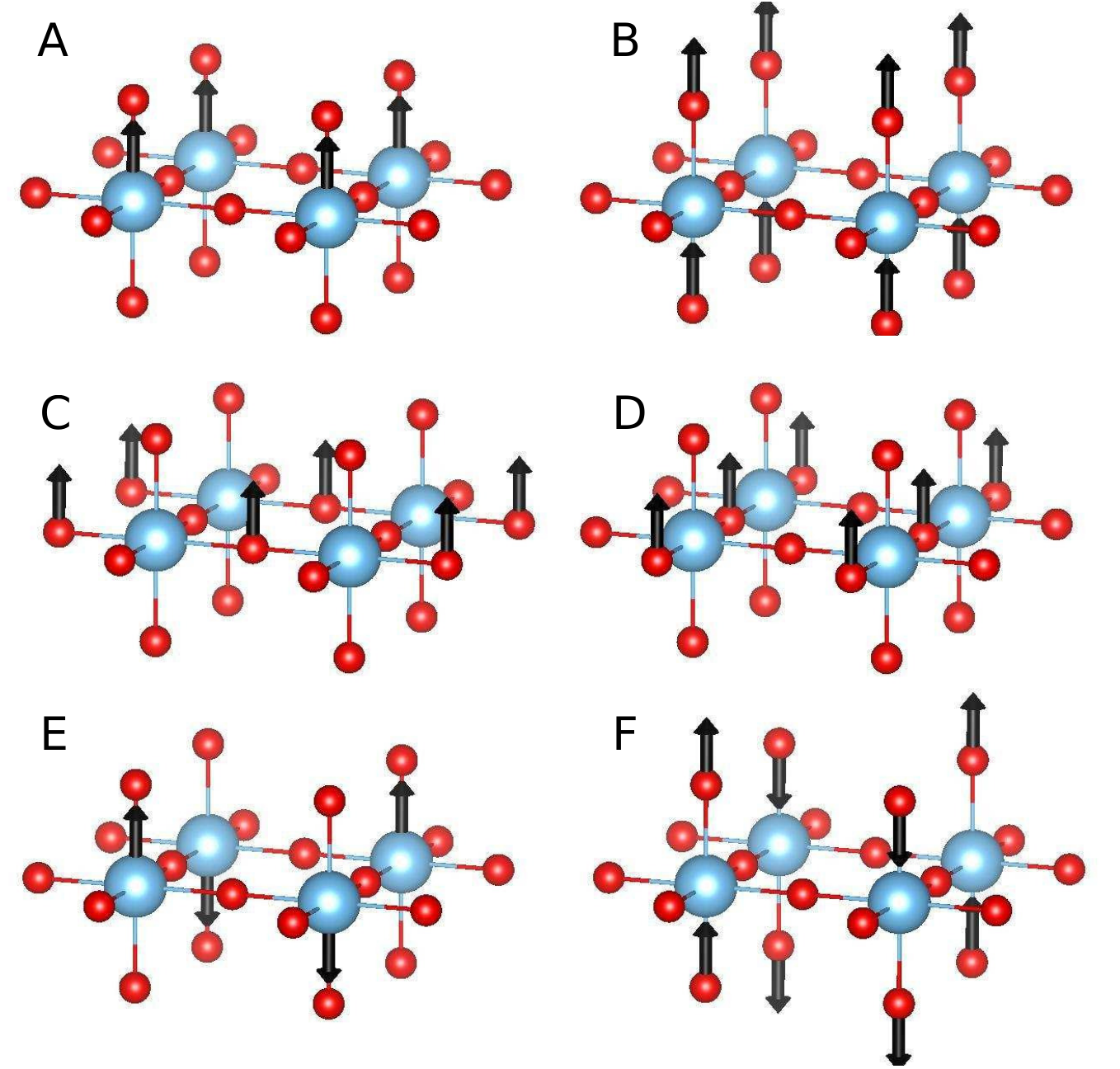}
  \caption{The atomic displacements we use to construct the force constants matrix. Panels A, B, C and D are the $\Gamma$ displacements of the Ti, O$_\parallel$ and two O$_\perp$ displacements respectively. We also displace the Ba atoms along a cubic axis, but only the Ti-O sublattice is shown for clarity. Panels E and F are the $M$-point displacements of Ti and O$_\parallel$. The blue spheres represent Ti atoms whereas the red spheres represent O atoms.}
  \label{fig:phonons}
\end{figure}

\subsubsection{\label{fevsafe}Driving force for ferroelectric versus antiferroelectric ordering.}
Given a collection of local dipoles, classical electrostatics tells us that a parallel alignment of dipoles is preferred energetically in the longitudinal direction, but an antiparallel alignment is preferred in the transverse direction. What this means in a real material is that if there was no other interaction than the long-range Coulomb interaction, antiferroelectricity is favored at least as much as ferroelectricity. Previous theory\cite{ghosez98, ghosez99} has shown that the cubic phase of BaTiO$_3$ is unstable to antiferroelectric as well as ferroelectric distortions. \emph{Why then does BaTiO$_3$ end up as ferroelectric, instead of an antiferroelectric?} We first answer this question for the insulating compound, and in the next section consider how the picture changes with electron doping. The zone center ($\Gamma$) lattice dynamical properties of BaTiO$_3$ in the cubic $Pm\bar{3}m$ phase are investigated by calculating the interatomic force constants matrix (FCM). The elements of the FCM are the second derivatives of the total energy with respect to atomic displacements. Diagonal elements -- known as self-force constants -- correspond to the second derivative with respect to the displacement of a given atom $u_i$, $\partial^2E/\partial u_i^2$, and are a measure of the energy cost (or gain) of displacing that atom. The off-diagonal elements correspond to derivatives with respect to the displacements of two different atoms $u_i$ and $u_j$, $\partial^2E/\partial u_i \partial u_j$, or the same atom in two different directions, and are measures of the strength of interaction between those atoms, \textit{i.e.,} the force constant of a virtual spring that couples the displacements of those atoms. In this study, we do not work with single atomic displacements, but rather the FCM for a certain $\vec{k}$ vector. For example, when we consider the zone center ($\Gamma$ point) FCM, the displacement of the Ba atom corresponds to the in-phase displacements of all the Ba atoms in every unit cell of an infinite crystal.  The FCM for cubic BaTiO$_3$ is a 15 $\times$ 15 block diagonal matrix that consists of three identical 5 $\times$ 5 blocks. Each block corresponds to displacements of the atoms along each of the three equivalent cubic axes. We calculate the FCM by displacing the atoms along one of the cubic axes in the patterns shown in Figure \ref{fig:phonons}A -- D. Note that there are two kinds of oxygen displacements: parallel with the Ti-O bond (O$_\parallel$, Figure \ref{fig:phonons}B) and perpendicular to the Ti-O bond (O$_\perp$, Figure \ref{fig:phonons}C and D). The FCM for the $\Gamma$ point is
\begin{equation}
C_\Gamma =
\left(
\begin{matrix}
6.59	&	\colorbox{blue!20}{-3.50}&	-1.25	&	-0.92	&	-0.92	\\
\colorbox{blue!20}{-3.49}	&	\colorbox{red!20}{-0.09}	&	\colorbox{red!20}{4.60}	&	-0.51	&	-0.51	\\
-1.25	&	\colorbox{red!20}{4.62}	&	\colorbox{red!20}{5.80}	&	\colorbox{blue!20}{-4.58}	&	\colorbox{blue!20}{-4.58}	\\
-0.92	&	-0.51	&	\colorbox{blue!20}{-4.59}	&	5.41	&	0.60	\\
-0.92	&	-0.51	&	\colorbox{blue!20}{-4.58	}&	0.60	&	5.41	
\end{matrix}
\right)
\label{eq:BTO_Gamma}
\end{equation}
in units of eV/\AA$^2$. The diagonal elements correspond to the self-force constants of the Ba, Ti, O$_\parallel$ and two O$_\perp$ displacements, from top left to bottom right. The self-force constant of the Ti atom (-0.09 eV/\AA$^2$) is the only negative one, in other words, just displacing the Ti atom by itself leads to an energy gain -- albeit a very small one -- whereas displacing any other atom by itself has an energy cost. In order to find the displacement pattern of the energy lowering ferroelectric (polar) mode, we diagonalize $C_\Gamma$ and look at its eigenvalues (force constants) and eigenvectors (referred to as eigenmodes or modes below). The ferroelectric mode has a negative force constant (-3.8 eV/\AA$^2$) and is dominated by O$_\parallel$ and Ti atom displacements, as expected given the charge transfer interaction between Ti and O. In fact, if we consider just the 2$\times$2 block corresponding to the Ti and O$_\parallel$ displacements (shown in red in Equation \ref{eq:BTO_Gamma}), the force constant of the ferroelectric mode is still negative and equal to -2.6 eV/\AA$^2$. In other words, if we clamp all the atoms and allow only Ti atom and O$_\parallel$ displacements, then BaTiO$_3$ still exhibits a ferroelectric instability. Even though the Ti displacements by themselves are not sufficient to significantly lower the energy, a coherent displacement of Ti and O$_\parallel$ can do so. However, the small contributions from Ba and O$_\perp$ ions have a nonzero contribution to the energy gain as well. This is due to the large off-diagonal matrix elements (shown in blue in Equation \ref{eq:BTO_Gamma}), which couple the displacements of these atoms with the Ti and O$_\parallel$ displacements. 

As mentioned above, previous first-principles calculations have established that in addition to a ferroelectric instability, BaTiO$_3$ also exhibits antiferroelectric instabilities at various points throughout the Brilloiun zone.\cite{ghosez98, ghosez99} For example, the force constants matrix at the $M$ point ($\vec{k}=(\pi/a,\pi/a,0)$ in Cartesian coordinates) is 
\begin{equation}
C_M =
\left(
\begin{matrix}
6.17	&	\colorbox{blue!20}{0.00}	&	0.00	&	0.00	&	0.00	\\
\colorbox{blue!20}{0.00}	&	\colorbox{red!20}{-0.03}	&	\colorbox{red!20}{4.39}	&	0.00	&	0.00	\\
0.00	&	\colorbox{red!20}{4.41}	&	\colorbox{red!20}{6.33}	&	\colorbox{blue!20}{0.00}	&	\colorbox{blue!20}{0.00}	\\
0.00	&	0.00	&	\colorbox{blue!20}{0.00}	&	7.29	&	0.00	\\
0.00	&	0.00	&	\colorbox{blue!20}{0.00}	&	0.00	&	7.29	
\end{matrix}
\right).
\label{eq:BTO_M}
\end{equation}
We used the same set of atomic displacements to compute $C_M$, but the $\vec{k}$ vector of the displacements is different, for example, see Figure \ref{fig:phonons}E-F. Even though the form of $C_M$ is different to $C_\Gamma$, the $2 \times 2$ block corresponding to Ti and O$_\parallel$ (highlighted again in red) is almost identical. If we again clamp all the atoms and allow only Ti atom and and O$_\parallel$ displacements, the force constant of the antiferroelectric mode is -2.3 eV/\AA$^2$, very similar to the $\Gamma$ point value. \emph{The driving force for the ferroelectric instability is thus the same as that for the antiferroelectric instability, that is, hybridization between the Ti $d$ and O $p$ states.} If it were only for the Ti and the O$_\parallel$ atoms, the strength of the ferroelectric and antiferroelectric instabilities would be very similar. However, $C_M$ does not have the large off-diagonal elements that $C_\Gamma$ has (due to symmetry), and as a result the antiferroelectric displacement of the Ti and O$_\parallel$ atoms does not couple with the displacement of the Ba and O$_\perp$ atoms. Hence, the reason that the ferroelectric instability is stronger is because the polar eigenmode contains a contribution from O$_\perp$ and Ba (a few percent each), which leads to an additional energy gain. It is this coupling between the Ti and O$_\parallel$ and O$_\perp$ and Ba displacements that renders the ferroelectric instability stronger than the antiferroelectric ones.

The preceding discussion has established that short-range interactions between the Ba and Ti cations play a significant role in driving ferroelectricity in BaTiO$_3$. However, there is also a contribution from a long-range Coulomb or electrostatic force. Strictly speaking, there are actually two types of long-range electrostatic interactions in an ionic crystal.\cite{maraddudin71} The first one is due to the macroscopic electric field created by a polar displacement and this is responsible for the splitting in frequency between the longitudinal optic and transverse optic phonons in ionic crystals; we ignore this interaction here as it can be easily `zeroed out' in both theory and experiment. The second kind of long-range electrostatic interaction, which we now consider, is known as the dipole-dipole interaction.\cite{ghosez96,ghosez98} Despite the name, this interaction has nothing to do with the classical dipole defined in Equation \ref{dip}. The dipole-dipole interaction makes a contribution to the force constants that is directly proportional to the Born effective charges of the atoms. The Born effective charge for a particular atom, $Z^\ast$, is defined as,
\begin{equation}
Z^\ast_{\alpha\beta} = \Omega\frac{\partial P_\beta}{\partial \tau_{\alpha}},
\label{bec}
\end{equation}
where $\Omega$ is the unit cell volume, $P$ is the polarization, $\tau$ is an atomic displacement and $\alpha$ and $\beta$ label different Cartesian directions. The Born effective charge is a tensor and can be interpreted as the ``amount of charge that effectively contributes to the polarization during the displacement''\cite{spaldin12} of a given atom, or the change in covalency with respect to the displacement of a given atom. It is important to keep in mind that the Born effective charge thus has a different physical meaning than the formal charge. In addition, in contrast to formal charges, Born effective charges are rigorously defined and experimentally measurable. The Born effective charges of the Ti atom and the O$_\parallel$ component of the O atom in BaTiO$_3$ are anomalously large compared to their formal values: +7.3 and -5.7, respectively.\cite{ghosez98b} This is to be expected given the charge transfer interaction between these atoms and their role in the ferroelectricity of BaTiO$_3$. Hence, the dipole-dipole interaction makes a significant contribution to the force constants of these atoms. There are thus two contributions to the ferroelectricity in BaTiO$_3$: local interactions between particular atoms, which are short range, and a longer range component due to the dipole-dipole interaction.

\subsubsection{Doping Dependence of Polar Instabilities.}
We now consider how the FCM evolves under electron doping to elucidate the effect of charge carriers on ferroelectricity in not only BaTiO$_3$, but also CaTiO$_3$. The most stable structure of CaTiO$_3$ is not ferroelectric or even polar, however CaTiO$_3$ in the cubic $Pm\bar{3}m$ phase does exhibit a zone-center polar instability that is almost of the same magnitude as that in BaTiO$_3$. In contrast to BaTiO$_3$ however, previous theory\cite{benedek13} has shown that although the Ti atom does make a significant contribution to the polar eigenmode of CaTiO$_3$, it is actually dominated by Ca displacements. As we show below, this difference is key to the persistence of the polar instability in CaTiO$_3$ under electron doping.

\begin{figure}[h]
\centering
  \includegraphics[width=0.30\textwidth]{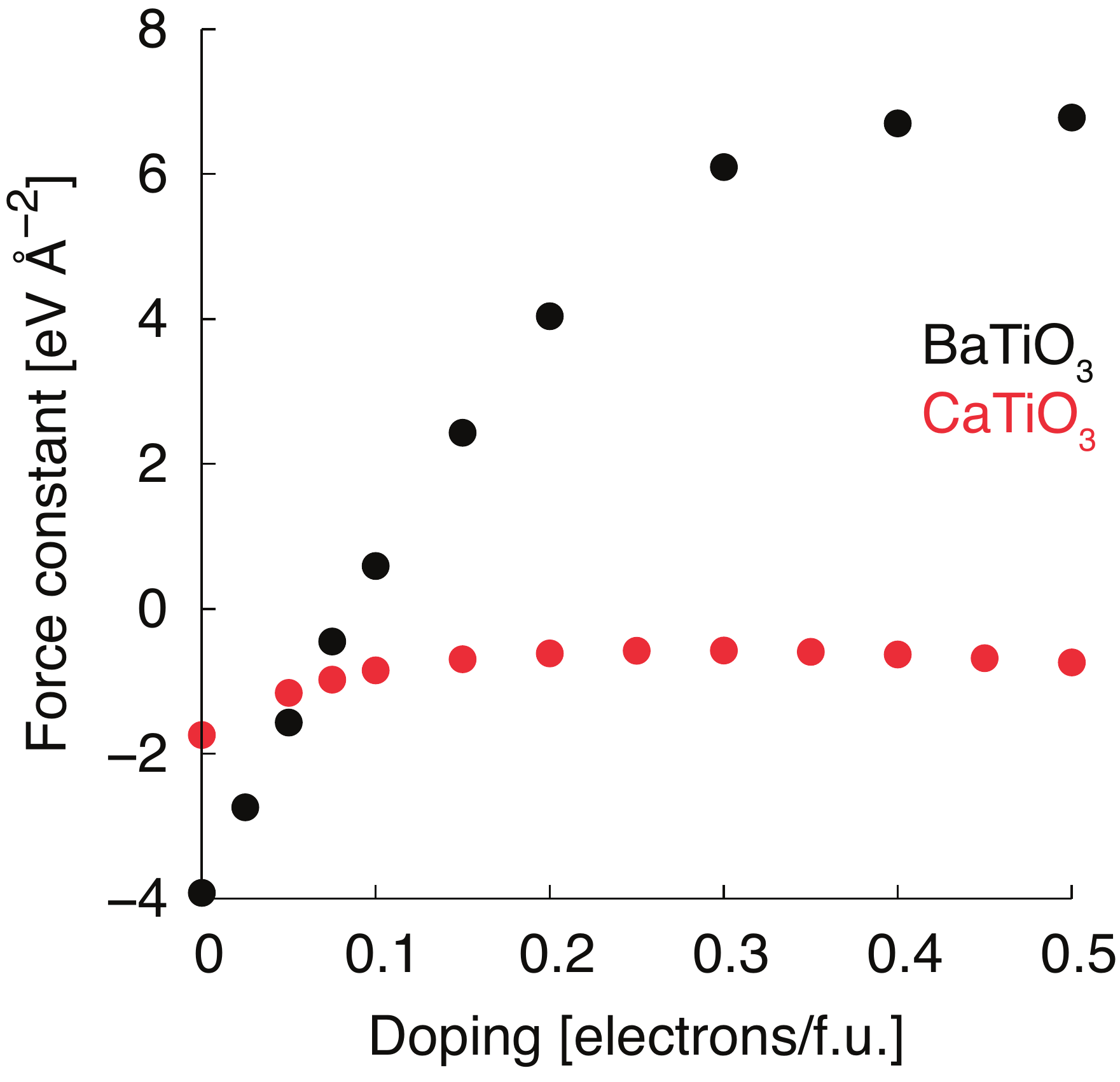}
  \caption{Variation in the force constant of the ferroelectric mode of BaTiO$_3$ and CaTiO$_3$ as a function of electron doping.}
  \label{fig:FE_Doped}
\end{figure}

The black data points in Figure \ref{fig:FE_Doped} show how the force constant of the ferroelectric mode of BaTiO$_3$ evolves with electron doping. Consistent with previous studies,\cite{hwang10, kolodiazhnyi10, wang12} even a small amount of doping (between 0.05 and 0.10 electrons/f.u.) suppresses the ferroelectric instability, \textit{i.e.} the force constant becomes positive, indicating that a ferroelectric distortion no longer lowers the energy of the cubic phase. The FCM for the system that is doped with 0.20 electrons/f.u. is 
\begin{equation}
C_{0.20} =
\left( 
\begin{matrix}
6.27	&	\colorbox{blue!20}{-3.30}	&	-1.17	&	-0.90	&	-0.90	\\
\colorbox{blue!20}{-3.29}	&	\colorbox{red!20}{5.47}	&	\colorbox{red!20}{0.89}	&	-1.53	&	-1.53	\\
-1.17	&	\colorbox{red!20}{0.89}	&	\colorbox{red!20}{8.62}	&	\colorbox{blue!20}{-4.17}	&	\colorbox{blue!20}{-4.17}	\\
-0.89	&	-1.51	&	\colorbox{blue!20}{-4.19}	&	5.96	&	0.63	\\
-0.90	&	-1.55	&	\colorbox{blue!20}{-4.16	}&	0.63	&	5.98	
\end{matrix}
\right).
\label{eq:BTO_Gamma_Doped}
\end{equation}
Comparing this matrix with that for the undoped case (Eq. \ref{eq:BTO_Gamma}), we see that the greatest change is in the self-force constant of Ti -- it not only changes sign, but also its absolute value increases more than 50-fold under doping! A secondary effect of doping is a change in the character of the
polar eigenmode. While it consists of 60\% Ti displacements in undoped BaTiO$_3$, the increase in the Ti self-force constant decreases the Ti contribution to the mode. A doping level of 0.20 electrons/f.u. decreases the Ti contribution to the eigenvector to only 41\%, while raising the Ba contribution to 18\%. In other words,
\emph{doping changes the character of the ferroelectric mode from almost purely B-site to mixed A- and B-site.} We will return to this point below.

Moving now to CaTiO$_3$, we mentioned above that the most stable structure for this material is not polar. At room temperature, CaTiO$_3$ adopts an orthorhombic structure with $Pnma$ symmetry, which is characterized by rotations of the TiO$_6$ octahedra. The rotations serve to optimize the coordination environment of the Ca cation, which is underbonded in the $Pm\bar{3}m$ phase. However, CaTiO$_3$ does have a ferroelectric instability in the cubic structure, which is driven not by charge transfer between the Ti and O atoms, but by the local bonding preferences of the Ca cation. Does this make the polar instability of CaTiO$_3$ more resistant against charge carriers?

We calculate the FCM for CaTiO$_3$ at the $\Gamma$ point using the same set of atomic displacements as for BaTiO$_3$: 
\begin{equation}
C_\Gamma =
\left(
\begin{matrix}
\colorbox{green!20}{0.87}	&	-2.46	&	-1.03	&	1.31	&	1.31	\\
-2.46	&	\colorbox{red!20}{3.88}	&	\colorbox{red!20}{-3.37}	&	0.97	&	0.97	\\
-1.03	&	\colorbox{red!20}{-3.34}	&	\colorbox{red!20}{16.19}	&	-5.91	&	-5.91	\\
1.32	&	0.99	&	-5.91	&	\colorbox{blue!20}{2.42}	&	1.18	\\
1.32	&	0.99	&	-5.91	&	1.18	&	\colorbox{blue!20}{2.42}
\end{matrix}
\right).\label{eq:CTO_Gamma}
\end{equation}

The lowest eigenvalue of this matrix is -1.74 eV/\AA$^2$, corresponding to a ferroelectric instability. The smallest self-force constant belongs to the Ca cation (0.87 eV/\AA$^2$, green box), and it is followed by the O$_\perp$ displacements (highlighted in blue), as expected for an A-site driven ferroelectric. Although the self-force constant of the Ti atom is quite high (3.88 eV/\AA$^2$), Ti displacements help to lower the total energy (the polar mode has a 12\% Ti contribution). Nevertheless, the polar eigenmode is dominated by displacements of the Ca cation and and O$_\perp$ ions (85\% in total). 

The red data points in Figure \ref{fig:FE_Doped} show the force constant of the ferroelectric mode as a function of electron doping. Like BaTiO$_3$, the force constant increases (becomes less negative) with doping, but unlike BaTiO$_3$, the force constant plateaus at a negative value for increased doping. This is despite the fact that the change in the form of the FCM as a function doping is similar to that in BaTiO$_3$. The FCM's for 15\% and 30\% electron doping are as follows: 
\begin{equation}
C_{0.15} =
\left(
\begin{matrix}
\colorbox{green!20}{0.64}	&	-2.34	&	-0.96	&	1.33	&	1.33	\\
-2.34	&	\colorbox{red!20}{9.10}&	\colorbox{red!20}{-6.59}	&	-0.09	&	-0.09	\\
-0.95	&	\colorbox{red!20}{-6.58}	&	\colorbox{red!20}{18.49}	&	-5.48	&	-5.48	\\
1.33	&	-0.03	&	-5.51	&	\colorbox{blue!20}{2.99}	&	1.21	\\
1.33	&	-0.11	&	-5.47	&	1.21	&	\colorbox{blue!20}{3.04}	
\end{matrix}
\right).
\label{eq:CTO_Gamma_Doped_15}
\end{equation}
\begin{equation}
C_{0.30} =
\left(
\begin{matrix}
\colorbox{green!20}{0.41}	&	-2.23	&	-0.86	&	1.34	&	1.34	\\
-2.23	&	\colorbox{red!20}{13.31}	&	\colorbox{red!20}{-9.33}	&	-0.87	&	-0.87	\\
-0.85	&	\colorbox{red!20}{-9.33}	&	\colorbox{red!20}{20.43}	&	-5.12	&	-5.12	\\
1.34	&	-0.85	&	-5.13	&	\colorbox{blue!20}{3.43}	&	1.21	\\
1.34	&	-0.85	&	-5.13	&	1.21	&	\colorbox{blue!20}{3.43}	
\end{matrix}
\right).
\label{eq:CTO_Gamma_Doped_30}
\end{equation}
Just as in BaTiO$_3$, the most dramatic change in the FCM of CaTiO$_3$ under doping is in the $2 \times 2$ block corresponding to the Ti and O$_\parallel$ displacements, shown in red. The hardening of these components decreases the absolute value of the force constant. However, the components that drive the ferroelectric instability, namely the Ca and the O$_\perp$ atoms (shown in green and blue, respectively) do not change as dramatically under doping, and in fact, the Ca self-force constant softens with doping. Similarly to BaTiO$_3$, the hardening of the Ti force constant under doping has the effect of decreasing the Ti contribution to the polar eigenmode of CaTiO$_3$. Doping reduces the contribution of the Ti atom from 12\% in the undoped case, to just 0.2\% in the case of doping with 0.20 electrons/f.u. Conversely, the combined contribution of Ca and O$_\parallel$ increases to 98\%. The presence of charge carriers thus hardens certain components of the FCM and suppresses ferroelectricity in BaTiO$_3$, whereas ferroelectricity persists under doping in CaTiO$_3$. How can we explain this difference?

Recall from our discussion in Section \ref{fevsafe} that there are two contributions to ferroelectricity in BaTiO$_3$, a short-range contribution and a longer range dipole-dipole contribution; it is instructive to consider what happens to each contribution under doping. We have just seen how doping stiffens the Ti component of the FCM for BaTiO$_3$. This means that the crystal cannot lower its energy through displacements of the Ti atoms, and since the displacements of the Ti atoms are coupled to the Ba and O$_\perp$ displacements, a reduced tendency for the Ti atoms to displace also results in a reduced tendency for Ba atom and O$_\perp$ displacements. The energy gain coming from the short-range contribution as a result of a polar distortion thus disappears under doping. In addition, the presence of charge carriers screens the dipole-dipole interaction and so the energy gain coming from that contribution also eventually disappears under doping. In contrast, for CaTiO$_3$ the local bonding contribution is much larger than for BaTiO$_3$ and the dipole-dipole contribution is much less important. As mentioned above, the Ca cation is significantly underbonded in the cubic perovskite structure and this is reflected in the large contribution of the Ca cation to the polar eigenmode of the cubic phase. The displacements of the Ca cation associated with the polar eigenmode are not the result of an SOJT distortion, but are instead due to local electrostatic/ionic size mismatch effects. Indeed, the Ca cation in CaTiO$_3$ has a formal valence of +2 and so the 4$s$ states are empty and lie above the Ti 3$d$ states in the conduction band; the 3$s$ and 3$p$ states of Ca are filled and are located deep within the valence band, well below the Fermi energy. Turning now to the dipole-dipole contribution, the Born effective charge of the Ca atom in undoped cubic CaTiO$_3$ is +2.6, close to its formal valence. The dipole-dipole interaction thus makes a much smaller contribution to the Ca force constant than it does to the Ti and O$_\parallel$ force constants of BaTiO$_3$. In other words, charge carriers may screen the dipole-dipole interaction in the doped material and stiffen the Ca force constant somewhat, but the local bonding contribution is essentially resistant to doping. \emph{To summarize, differences in the ferroelectric mechanisms of the two materials means that long-range electrostatic forces (what we have been calling the dipole-dipole interaction) are essential to the emergence of a polar structure in BaTiO$_3$, but are of only secondary importance in CaTiO$_3$.}

\begin{figure}
\centering
\includegraphics[width=8cm]{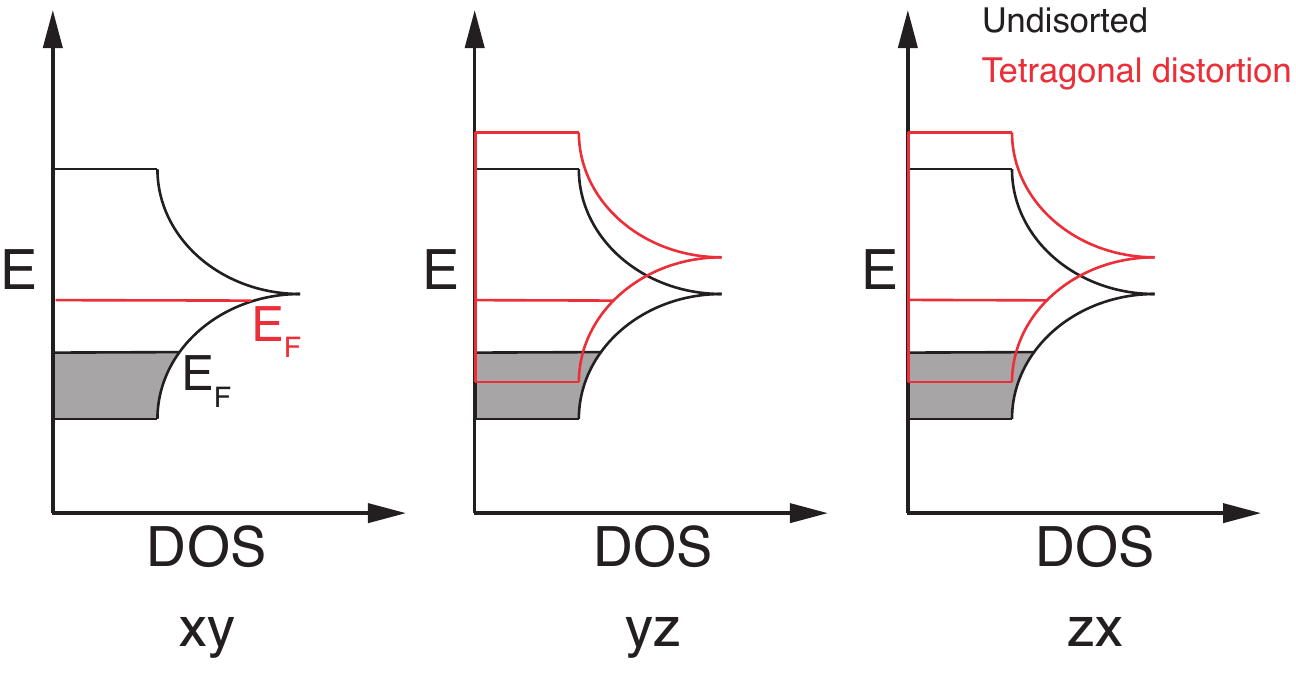}
\caption{\label{sojt}Schematic showing the Fermi level (E$_F$) and energy levels of the Ti $t_{2g}$ orbitals in cubic and distorted BaTiO$_3$ with a non-zero concentration of $d$ electrons. The shading represents occupied states in the conduction band of the doped material.}
\end{figure}

There is another way to think about the suppression of ferroelectricity with doping in BaTiO$_3$, or the stiffening of the Ti and O$_\parallel$ force constants of both BaTiO$_3$ and CaTiO$_3$. Wheeler and co-workers\cite{wheeler86} developed a simple model based on H\"{u}ckel theory to explain structural distortions as a function of $d$ electron count in materials containing metal-centered, vertex-sharing polyhedra, including perovskites; we follow essentially their treatment here. This model neglects interactions between the electrons and simplifies the structural distortion such that only the Ti atom moves. It is nonetheless instructive and provides another perspective on the phenomena we have been discussing. In the undistorted and undoped cubic perovskite structure, the $t_{2g}$ $d$ orbitals of the Ti atom are degenerate and are located at the bottom of the conduction band. Now consider an SOJT distortion that lengthens the $c$ axis, which is parallel with the Cartesian $z$ direction, \textit{i.e.} a tetragonal distortion. The $xy$ orbital stays at roughly the same energy, whereas the $yz$ and $zx$ orbitals slightly shift up in energy, as shown in Figure \ref{sojt}. Conversely, the O $p$ orbitals, located at the top of the valence band, shift down in energy slightly. The energy shift of both sets of orbitals can be estimated from perturbation theory and is of the same magnitude,\cite{wheeler86} \textit{i.e.} the $yz$ and $zx$ orbitals move up by the same amount that the O $p$ orbitals move down. Hence, the distorted phase has a larger band gap than the undistorted one.

We now increase the electron concentration to $d^1$. Each of the $t_{2g}$ orbitals will now be $\frac{1}{6}$ filled (accounting for spins). When the crystal undergoes a distortion to the tetragonal phase, the $yz$ and $zx$ orbitals will again shift up in energy, and their electrons will flow to the $xy$ orbital. However, in order for the number of occupied states in the conduction band to remain constant (between the undistorted and distorted phases), the Fermi level must shift up, which will raise the total energy of the system. In the model of Wheeler, \textit{et al}, this energy penalty is not compensated by the energy lowering from the O $p$ bands being pushed down. Hence, as the $d$ electron count rises from zero, the driving force for the distortion disappears. Note that this model does not account for the persistence of a polar distortion under doping in CaTiO$_3$, because it considers only one mechanism (SOJT) as the driving force for the distortion.

We also considered the effects of doping on the polar instability in SrTiO$_3$ and found that it is suppressed essentially as soon as there is any occupation of the Ti $d$ levels. This is because the SOJT instability in SrTiO$_3$ is not as strong as it is BaTiO$_3$; the Sr atom is also larger and less underbonded in the cubic phase compared to CaTiO$_3$, and so the Sr atom does not make as much of a contribution to the polar eigenmode as the Ca atom does to the polar eigenmode of CaTiO$_3$. There is thus a crossover in the behavior of the titanate perovskites. BaTiO$_3$ is a B-site dominated ferroelectric in which Ti-O hybridization gives rise to a strong SOJT distortion that can persist under some electron doping. CaTiO$_3$ is an A-site dominated ferroelectric characterized by atomic displacements that are resistant to electron doping, because they do not depend on a cross-gap charge transfer mechanism that enhances the dipole-dipole contribution to their force constant, which is subsequently screened by electrons. SrTiO$_3$ does not fall squarely into either of these two categories, and so its ferroelectric instability is strongly suppressed under even very low levels of electron doping.

\subsection{Putting It All Together: Similarities Between LiOsO$_3$, Doped Titanate Perovskites and Other Polar Metals}
The mechanisms through which LiOsO$_3$ and CaTiO$_3$ transition to polar structures share a common feature: both appear to be driven by local ionic size mismatch effects. Such materials are known as `geometric' ferroelectrics\cite{vanaken04,ederer06} in the ferroelectrics literature (perhaps the most well-studied compound of this type is the multiferroic YMnO$_3$\cite{vanaken04,fennie05}). The $Pmc2_1$ ground state of the predicted polar metal SrCaRu$_2$O$_6$\cite{puggioni14} is also established through a related but somewhat different mechanism to LiOsO$_3$ and CaTiO$_3$, namely the trilinear coupling mechanism described in the Introduction. In the most general trilinear coupling case, two zone-boundary modes, $R_1$ and $R_2$, of different symmetries couple to a polar zone-center mode, $P$. The product of $R_1$ and $R_2$ establishes the polar space group symmetry, $P \sim R_1 \oplus R_2$. This is a rather general mechanism that can induce ferroelectricity in many families of layered oxides;\cite{withers91,snedden03,dawber08,goff09,knapp06b,king07,king10,fukushima11,rondinelli12} see Ref.\ \onlinecite{benedek15} for a recent review. In SrCaRu$_2$O$_6$ the two zone-boundary modes $R_1$ and $R_2$ are RuO$_6$ octahedral rotations. Such distortions optimize the A-site bonding environment (Sr and Ca here), which is underbonded in the undistorted non-polar phase. Most importantly, as we have already discussed, octahedral rotations are generally not associated with any kind of cross-gap hybridization and therefore do not change the electronic structure at the Fermi level. Ref. \onlinecite{puggioni14} explicitly shows using first-principles calculations that neither the octahedral rotations nor the polar mode significantly changes the density of states at the Fermi level. SrCaRu$_2$O$_6$, like both LiOsO$_3$ and CaTiO$_3$, would thus appear to be a geometric `ferroelectric'.\footnote{The difference between the mechanisms of SrCaRu$_2$O$_6$ and LiOsO$_3$ (and CaTiO$_3$) is that SrCaRu$_2$O$_6$ is classed as an improper `ferroelectric': the polar phase is established through a combination of two zone-boundary lattice modes. In contrast, the polar phases of LiOsO$_3$ and CaTiO$_3$ emerge via the condensation of a single zone-center polar lattice mode, and these materials are therefore proper `ferroelectrics'. All three materials however are geometric `ferroelectrics'. We emphasize that these materials are metals and they therefore cannot be actual ferroelectrics, \textit{i.e.} materials with a switchable polarization.} Although our results appear to suggest that polar metals must then be geometric `ferroelectrics', the story is somewhat more complicated, as discussed further below.
 
\subsection{Design Principles and Future Directions}
The two key conclusions to be drawn from our work so far are that for the materials we have considered, 1) there does not appear to be a fundamental incompatibility between polarity and metallicity, and 2) the polar phase of the materials resistant to doping or metallicity (CaTiO$_3$ and LiOsO$_3$) emerges through a geometric mechanism, rather than one involving charge transfer or hybridization. Hence, the most promising class of materials to search for new polar metals would appear to be metallic compounds with a tendency towards a polar distortion \emph{that emerges through a geometric mechanism}. The second conclusion is essentially a restatement of the ``weak coupling hypothesis'' formulated by Puggioni and Rondinelli,\cite{puggioni14} which states that, ``the existence of any non-centrosymmetric metal relies on weak coupling between the electrons at the Fermi level, and the (soft) phonon(s) responsible for removing inversion symmetry". The materials we have considered in this study certainly seem to satisfy the weak coupling hypothesis, however there are exceptions. Ref. \onlinecite{xiang14} studied the hypothetical compound TiGaO$_3$, which like LiOsO$_3$ also adopts a structure with $R3c$ symmetry as the lowest energy phase. Ti is in a nominally 3+ oxidation state with a $d^1$ valence electron configuration. An analysis of the density of states for TiGaO$_3$ showed that the Ti $d$ states dominated the density of states around the Fermi level. However, the mechanism through which the polar phase emerges from the non-polar $R\bar{3}c$ structure is the same as LiOsO$_3$, that is, the distortion is driven by the bonding preferences of the Ti atom. Hence, in TiGaO$_3$ the Ti atom is responsible for both the metallicity and the polar instability.      

Bennett and co-workers also recently showed\cite{bennett12b, bennett13} that there are a number of potential ferroelectrics among hexagonal LiGaGe-type intermetallic compounds; the considered materials are either already known in polar structures or are predicted from first-principles to be polar, though polarization switching has not been experimentally demonstrated for any insulating LiGaGe-type material. In this family of materials, the driving force for ferroelectricity (or antiferroelectricity) is the preference of the metal atoms for $sp^3$ bonding rather than planar $sp^2$ bonding. The atoms displaced to form $sp^3$ bonds buckle the atomic layers, and depending on the ordering of the bonds, the material becomes either ferroelectric (LiGaGe-type structure) or antiferroelectric (MgSrSi-type). Of the 18 non-rare earth compounds reported in the ICSD to have the non-centrosymmetric LiGaGe-type structure, only two are insulating. Theory predicts\cite{bennett12b} that a number of as-yet unsynthesized LiGaGe-type compounds should also be metallic. Interestingly, of these, many are further predicted to undergo a metal-insulator transition \emph{coincident with the polar one.} Some materials are insulating in their nonpolar phases and metallic after a polar distortion, and for other materials the opposite is true. This clearly suggests an interplay between the polar distortion and the electronic structure at the Fermi level.

How then should we think about the design of polar metals? Our work has illustrated a connection between the compatibility of metallicity and polar distortions and the mechanism through which the polar phase emerges (a connection that was also noticed by Ref. \onlinecite{puggioni14}). Although it is tempting to conclude that polar or non-centrosymmetric metals must be geometric `ferroelectrics', there exist several (and quite probably more) counter-examples, discussed above. Extensive investigations by different groups over the past several years have resulted in the discovery of new ferroelectric mechanisms, however as far as we are aware, much of this effort has focused on a select few families of materials, particularly complex oxides. There is likely much still to learn about mechanisms of polar distortion in other classes of materials, particularly the intermetallics, and the manner in which charge carriers may modify, compete with or suppress these distortions will probably differ compared to the oxides. Hence, we believe that the design process for new polar metals should start with an understanding of ferroelectric mechanisms in the materials family of interest.

\section{Conclusions}
In the Introduction we asked whether long-range electrostatic forces really are the driving force for polar distortions. Our work has shown that they can be in \emph{some} cases, though this statement requires qualification. First, it has not been uncommon to view polar distortions in an approximate way as simply the displacement of one type of atom, for example, the off-centering displacement of the Ti atom in the case of the titanate perovskites. We have shown for the titanates that it is critical to consider the full complexity of the polar eigenmode (the displacements of other atoms in the unit cell), since otherwise it is not possible to explain the preference for ferrroelectricity over antiferroelectricity. Secondly, our work has demonstrated that although electron doping suppresses the polar instability of BaTiO$_3$, `ferroelectricity' can persist in electron-doped CaTiO$_3$ and metallic LiOsO$_3$. In the case of BaTiO$_3$, the long-range dipole-dipole interaction, which we again emphasize has nothing to do with the local dipoles discussed earlier, makes a large contribution to the force constant of the Ti atom, and to the force constant of the O atoms when they are displaced parallel to the Ti-O bond. This contribution, which is screened by free electrons once they reach a critical concentration, arises from the hybridization between the Ti $d$ and O $p$ states and manifests as an anomalously large Born effective charge for these atoms. In addition, the short-range contribution to the polar instability is not strong enough in BaTiO$_3$ to compensate for the weakened dipole-dipole interaction. In contrast, the polar phases of both CaTiO$_3$ and LiOsO$_3$ emerge through a different, geometric mechanism and hence the short-range contribution (essentially resistant to charge carriers) to the force constants of the Ca and Li atoms is much larger than the dipole-dipole contribution. Hence, in these materials, the long-range electrostatic dipole-dipole interaction is not really the driving force for polar distortions. Finally, we have used the insights gained from our fundamental study on mechanisms of inversion-symmetry breaking in metallic systems to comment on design principles for new polar metals. Although extensive work on the ferroelectric mechanisms of complex oxides suggests that a fruitful approach may be to search for metals with a tendency towards a polar distortion through a geometric mechanism, other materials families may require a different approach. Indeed, understanding the fundamental chemical and physical factors that give rise to polar distortions has been the starting point for the design of various kinds of functional materials, including multiferroics, and materials with predicted electric-field controllable metal-insulator transitions and Jahn-Teller distortions;\cite{bristowe15,varignon15} the utility of polar distortions in enabling the design of materials with novel couplings or properties is truly remarkable. We hope our work both encourages further investigations of the mechanisms of inversion-symmetry breaking in metals, and aids in the design and exploration of the properties of these fascinating systems.  

\section{Acknowledgements}
T. B. was supported by the Rutgers Center for Materials Theory. The authors thank Guru Khalsa, Joseph Bennett, Craig Fennie and James Rondinelli for helpful discussions.

%%%REFERENCES%%%
\bibliography{benedek} %You need to replace "rsc" on this line with the name of your .bib file

\end{document}